\def\year{2021}\relax
\documentclass[letterpaper]{article} 
\usepackage{aaai21}  
\usepackage{times}  
\usepackage{helvet} 
\usepackage{courier}  
\usepackage[hyphens]{url}  
\usepackage{graphicx} 
\usepackage{natbib}  
\usepackage{caption} 
\usepackage{dirtytalk}
\usepackage{amsmath}
\usepackage{multirow}

\title{A Study of F0 Modification for X-Vector Based Speech Pseudonymization Across Gender}
\author {
    Pierre Champion,\textsuperscript{\rm 1}
    Denis Jouvet, \textsuperscript{\rm 2}
    Anthony Larcher \textsuperscript{\rm 1} \\
}
\affiliations {
    \textsuperscript{\rm 1} Université de Lorraine, CNRS, Inria, LORIA, F-54000 Nancy, France. \\
    \textsuperscript{\rm 2} Le Mans Université, LIUM, France \\
    \{pierre.champion, denis.jouvet\}@inria.fr, anthony.larcher@univ-lemans.fr
}

\begin{document}

\maketitle

\begin{abstract}
Speech pseudonymization aims at altering a speech signal to map the identifiable personal characteristics of a given speaker to another identity.  In other words, it aims to hide the source speaker identity while preserving the intelligibility of the spoken content. This study takes place in the VoicePrivacy 2020 challenge framework, where the baseline system performs pseudonymization by modifying x-vector information to match a target speaker while keeping the fundamental frequency (F0) unchanged. We propose to alter other paralinguistic features, here F0, and analyze the impact of this modification across gender.
We found that the proposed F0 modification always improves pseudonymization.
We observed that both source and target speaker genders affect the performance gain when modifying the F0.

\end{abstract}
\section{Introduction}
\label{sec:intro}

In many applications, such as virtual assistants, speech signal is sent from the 
device to centralized servers in which data is collected, processed, and stored. 
Recent regulations, e.g., the General Data Protection Regulation (GDPR) \cite{gdpr} in the EU, 
emphasize on privacy preservation and protection of personal data.
As speech data can reflect both biological and behavioral characteristics 
of the speaker, it is qualified as personal data \cite{nautschGDPRSpeechData2019}.
The research reported in this paper has been done in the context of the VoicePrivacy challenge framework \cite{tomashenkoVoicePrivacy2020Challenge}, which is one of the first attempt of the speech community to encourage research on this topic,
define the task, introduce metrics, 
datasets and protocols.

Anonymization is performed to suppress the personally identifiable
paralinguistic information from a speech utterance while maintaining the linguistic
content. The task of the VoicePrivacy challenge 
is to degrade automatic speaker verification performance, by removing speaker identity 
as much as possible, while keeping the linguistic content intelligible.
This task is also referred to as \textit{speaker 
anonymization}\cite{fangSpeakerAnonymizationUsing2019} 
or \textit{de-identification} \cite{magarinos2017reversible}.
\vspace{14mm}

Anonymization systems in the VoicePrivacy challenge should satisfy the following requirements:
\begin{itemize}
\item output a speech waveform;
\item conceal the speaker's identity;
\item keep the linguistic content intelligible;
\item modify the speech signal of a given speaker to always sound like a unique target pseudo-speaker, while different speaker's speech must not be similar.
\end{itemize}
The fourth requirement constraints the system to have a one-to-one mapping between 
the real speaker identities and a pseudo-speaker. Such system can be considered 
as a voice conversion system where the output speaker identity resides in a pseudonymized
space.

The GDPR defines pseudonymization as: \label{gdpr_def}
\textit{\say{processing of personal data in such a manner that the personal data can no longer be attributed to a specific data subject without the use of additional information, provided that such additional information is kept separately and is subject to technical and organizational measures to ensure that the personal data are not attributed to an identified or identifiable natural person}}(Art.4.5 of the GDPR \cite{gdpr}).
pseudonymization techniques differ from anonymization techniques. With anonymization, data is modified so that any information that may serve as an identifier to a subject is deleted. 
pseudonymization enhances privacy by replacing most identifying information within data by artificial identifiers. Per the requirements imposed by the VoicePrivacy challenge, and the above
definition from GDPR, the challenge imposes contestants to build pseudonymization systems.
The VoicePrivacy challenge focuses on modifying the speech characteristics; while keeping the linguistic content unchanged; hence removing personal information from the linguistic content is not part of that challenge.

Recently, Fang et al. \cite{fangSpeakerAnonymizationUsing2019} proposed a speech synthesis pipeline where only the continuous speaker representation (the x-vector \cite{snyder2018xvector}) is modified. Linguistic related information necessary to generate anonymized speech is left untouched.
The corresponding toolchain doesn't alter the fundamental frequency (F0) input values, and the articulation of speech sounds feature (the Phoneme Posterior-Grams (PPGs) \cite{ppgs}).

The F0 values of speech determine the perceived relative highness or lowness of the sound, it plays an indispensable role for the listener as it helps to perceive a variety of paralinguistic, and prosodic information \cite{PhonologyOfTone}.
Analysis of the F0, which is typically higher in female voices 
than in male voices, can be used to characterize speaker-related attributes.

In this paper, we use the pipeline proposed by Fang et al. \cite{fangSpeakerAnonymizationUsing2019} in the VoicePrivacy challenge 2020 \cite{tomashenkoVoicePrivacy2020Challenge}, and discuss what possible improvement may be obtained by modifying the F0 values.

The remainder of the paper is structured as follows. First, we review the baseline framework and explains the conversion process. Secondly we describes the experimental setup. Then we present and discuss the results. Finally, we concludes the paper.

\section{Anonymization technique} \label{anon_tech}
\subsection{The baseline system}
\label{sec:vcmodel}

The VoicePrivacy challenge provides two baseline systems: 
\textit{Baseline-1} that anonymizes speech utterances using x-vectors and neural waveform 
models \cite{fangSpeakerAnonymizationUsing2019} and \textit{Baseline-2} that performs 
anonymization using McAdams coefficient \cite{mcadams}. Our contributions are based on 
\textit{Baseline-1} which is referred to as the \textit{baseline} system in 
this paper.

\begin{figure}[h]
  \centering
  \includegraphics[width=1.00\linewidth]{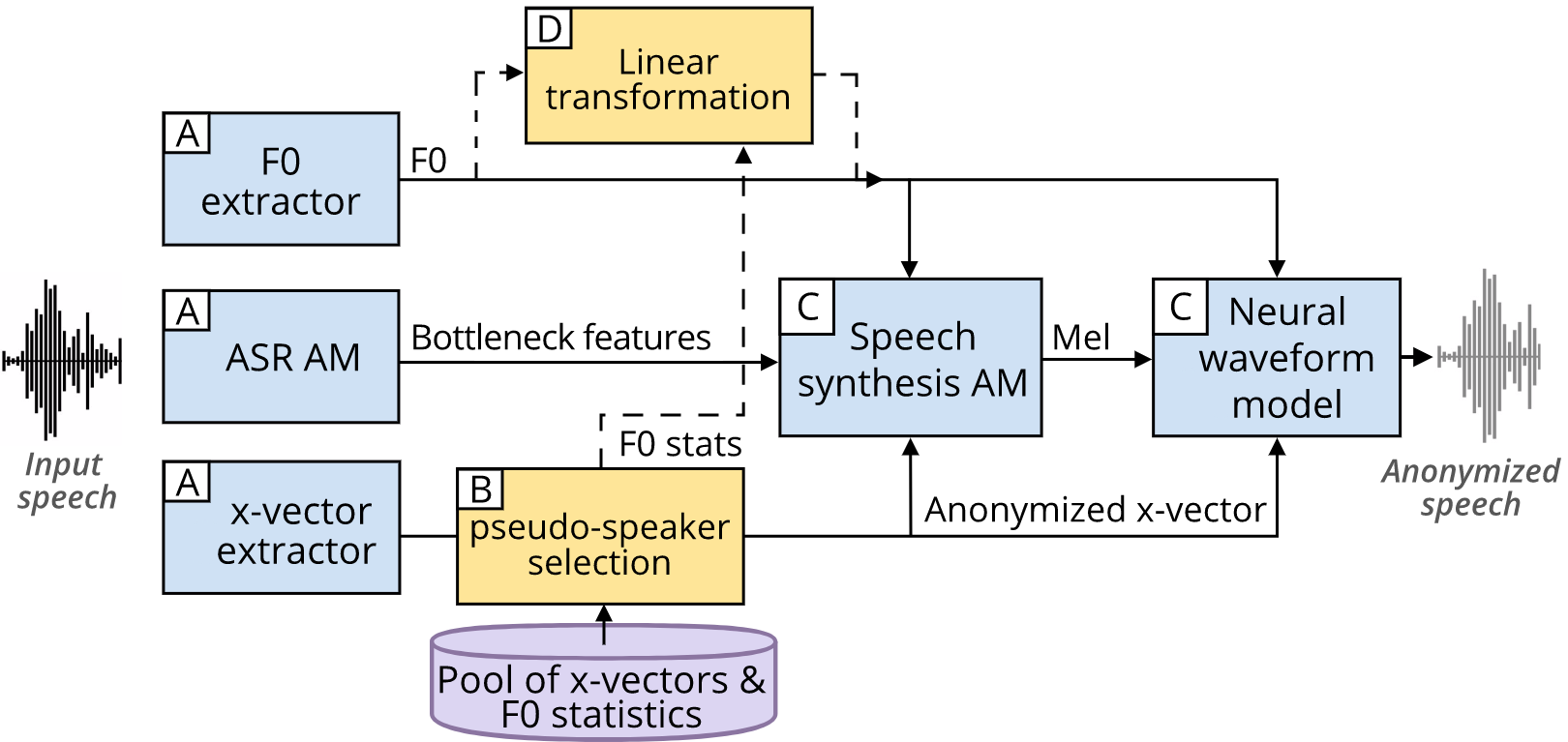}
  \caption{The speaker anonymization pipeline. Modules A, B and C are parts of the baseline model. We added module D to modify the F0 values, which are later used by modules C.}
  \label{fig:baseline.png}
\end{figure}

The central concept of the baseline system introduced in \cite{fangSpeakerAnonymizationUsing2019} 
is to separate speaker identity and linguistic content from an input speech utterance. 
Assuming that those information are disentangled, an anonymized speech waveform can be obtained by altering only the features that encode the speaker's identity.
The anonymization system illustrated in {Figure \ref{fig:baseline.png}} breaks down the 
anonymization process into three groups of modules: 
\textit{A - Feature extraction} comprises three modules that respectively extract fundamental frequency, PPGs like bottleneck features, and the speaker's x-vector from the input signal. Then, \textit{B - Anonymization} derives a new pseudo-speaker identity using knowledge gleaned from a pool of external speakers. Finally, \textit{C - Speech synthesis} synthesizes a speech waveform from the pseudo-speaker x-vector together with the original PPGs features, and the \textbf{original F0} using an acoustic model 
\cite{tomashenkoVoicePrivacy2020Challenge}
and a neural waveform model \cite{nsf}.
For all utterances of a given speaker, a single target pseudo-speaker is used to modify the input speech. This strategy, described as \textit{perm} in \cite{EvaluatingVoiceConversionbased2019}, ensures that a one-to-one mapping exists between the source speaker identity and the target pseudo-speaker.

\subsection{x-vector pseudonymization} \label{lab:x-vector}
Given the baseline system, where only the x-vector identity is changed, the selection algorithm used to derive a pseudo-identity plays an important role. 
Many criteria can be chosen to select the target pseudo-speaker identity.
Recent research made by \cite{brij_vpc_design} has outline multiple selection techniques for the VoicePrivacy Challenge. 
The baseline's pseudo-speaker selection is performed by averaging a set of x-vectors candidates from the speaker pool.
The candidate x-vectors are selected by retrieving the 200 furthest speakers given the original x-vector. From this subset of 200 x-vectors, a set of 100 x-vectors is randomly chosen to create the pseudo-speaker x-vector. 
Speaker's distances are queried according to the probabilistic linear discriminant analysis (PLDA).
The speaker pool is composed of speakers from the LibriTTS-train-other-500 \cite{libritts} dataset. This dataset is not used elsewhere in our experiments.

\subsection{Gender selection}
Information conveyed by the x-vector embeddings can be used for other tasks than speaker recognition/verification. Work by \cite{prob_x-vector} has shown that session and gender information, along with other characteristics, are also encoded in x-vectors.

The aforementioned x-vector anonymization procedure is designed to select a pseudo-speaker identity from the same gender as the source speaker.
Constraining the x-vector anonymization procedure to target x-vectors from same gender as the source is referred to as \textit{Same}, While constraining the selection to target the opposite gender is referred to as \textit{Opposite}.
\textit{Same}, and \textit{Opposite} gender selection were experimentally studied by \cite{brij_vpc_design}. 
Work on \textit{gender independent} selection still needs to be done.

In this paper, we focus our experience on \textit{Same} and \textit{Opposite} gender selections. We discuss the impact that F0 modification has on female and male speakers when using these two selection algorithms.

\subsection{Speech synthesis}
The speech synthesizer (cf. pipeline C in Figure \ref{fig:baseline.png}) in the
the VoicePrivacy baseline system is composed of a speech synthesis acoustic model, used to generate mel-fbanks features; and a vocoder, used to generate a speech signal.
The vocoder used in the baseline is a Neural Source-Filter (NSF) Waveform model \cite{nsf}. NSF models uses the F0 information to produce a sine-based excitation signal that is later transformed by filters into a waveform. Manipulating the F0 values will impact both the speech synthesis acoustic model and vocoder models to transform the speech signal.



\subsection{F0 modification}
In the VoicePrivacy baseline, the F0 values extracted from the source speech are directly used (unchanged) by the speech synthesizer pipeline (acoustic model and neural vocoder), even though a different target pseudo-speaker was selected. Multiple works have investigated F0 conditioned voice conversion \cite{F0_Bahmaninezhad2018ConvolutionalNN, f0_Huang2020UnsupervisedRD, F0_Qian2020F0ConsistentMN, F0_IndividualityPreservingSM}.
In some papers modifying the F0 improves the quality of the converted voice.
Motivated by those results, we propose to modify the F0 values of a source utterance from a given speaker (cf. module D in Figure \ref{fig:baseline.png}) by using the following linear transformation:
\begin{equation*}
    \hat{x}_{t}=\mu_{y}+\frac{\sigma_{y}}{\sigma_{x}}\left(x_{t}-\mu_{x}\right)
\end{equation*}
where $x_{t}$ represents the log-scaled F0 of the source speaker at the frame $t$, $\mu_{x}$ and $\sigma_{x}$ represent the mean and standard deviation for the source speaker. $\mu_{y}$ and $\sigma_{y}$ represents the mean and standard deviation of the log-scaled F0 for the pseudo-speaker.
The linear transformation and statistical calculation are only performed on voiced frames.
The mean and standard deviation for the target pseudo speaker are calculated by averaging information from the same 100 speakers selected to derive the pseudo-speaker x-vector.



\section{Experimental setup} \label{expe_setup}

\subsection{Data}
All experiments where based on the challenge publicly available  baseline\footnote{\url{https://github.com/Voice-Privacy-Challenge}}.
The development and evaluation sets are built from LibriSpeech \textit{test-clean}.
The pool of external speakers on which x-vectors and F0 statistics are computed is LibriSpeech \textit{train-other-500}.
Additional information on the number of speakers, and the gender distributions can be found in the evaluation plan \cite{tomashenkoVoicePrivacy2020Challenge}. 

\subsection{Attack models}
One of the requirements of the VoicePrivacy challenge is to \textit{conceal the speaker's identity}. To assess the robustness of anonymization systems, two attack models were designed (cf. evaluation plan).
The first scenario consists of a user who publishes
anonymized speech
and an attacker who uses one enrollment utterance of non-anonymized (original) speech to compute a linkability score. In this scenario (referred as \textbf{o-a} in {Figure \ref{fig:ASV-res}}), the goal is to ensure the \textbf{o}riginal speaker identity is not  the same as the one in the generated \textbf{a}nonymized speech. Performant systems are expected to show low linkability.
The second scenario consists of a user who also publishes anonymized speech, but this time, the attacker has itself anonymized an enrollment utterance using the same exact anonymization pipeline except for the random seed. This scenario (referred as \textbf{a-a} in {Figure \ref{fig:ASV-res}}) is defined as a \textit{Semi-Informed} attacker in work done by \cite{EvaluatingVoiceConversionbased2019}. \textbf{Hence, the pseudo-speaker corresponding to a given speaker in the enrollment set is different from the pseudo-speaker corresponding to that same speaker in the trial set, as mentioned in Section 3.3 of the evaluation plan}. Consequently, we also expect to have low linkability in this \textbf{a-a} scenario even through the attacker has gained some knowledge about the anonymization system.

\subsection{Utility and linkability metrics} \label{metric}
To evaluate the performance of the system in both linkability (\textit{speaker's concealing} capability) and utility (\textit{content intelligibility}) two systems are used. 
To assess the linkability, a pre-trained x-vector-PLDA based Automatic Speaker Verification (ASV) system provided by the challenge organizers is used. The privacy protection is measured in terms of $\text{C}_{llr}^{min}$ as this measure provides an application-independent \cite{cllr} evaluation score. As the Equal Error Rate (EER) measure is more often used in speaker verification, we present our result in terms of both EER and $\text{C}_{llr}^{min}$. Those metrics are computed using the cllr toolkit\footnote{\url{https://gitlab.eurecom.fr/nautsch/cllr/}} of the challenge.
For the utility, a pre-trained Automatic Speech Recognition (ASR) system provided by the challenge organizers is used to decode the anonymized speech and compute the Word Error Rate (WER$_\%$). In this challenge, the WER$_\%$ measure is used to evaluate how the content is kept intelligible.
Both ASR and ASV systems are trained on LibriSpeech \textit{train-clean-360} using Kaldi \cite{KaldiPovey}.
The higher the EER/$\text{C}_{llr}^{min}$, the better the systems are capable of \textit{\say{concealing a speaker identity}}. The lower the WER$_\%$ is, the more intelligible the anonymized speech is.

\section{Experimental results} \label{expe_results}

\begin{figure}[!ht]
  \centering
  \centerline{\includegraphics[width=1.0\linewidth]{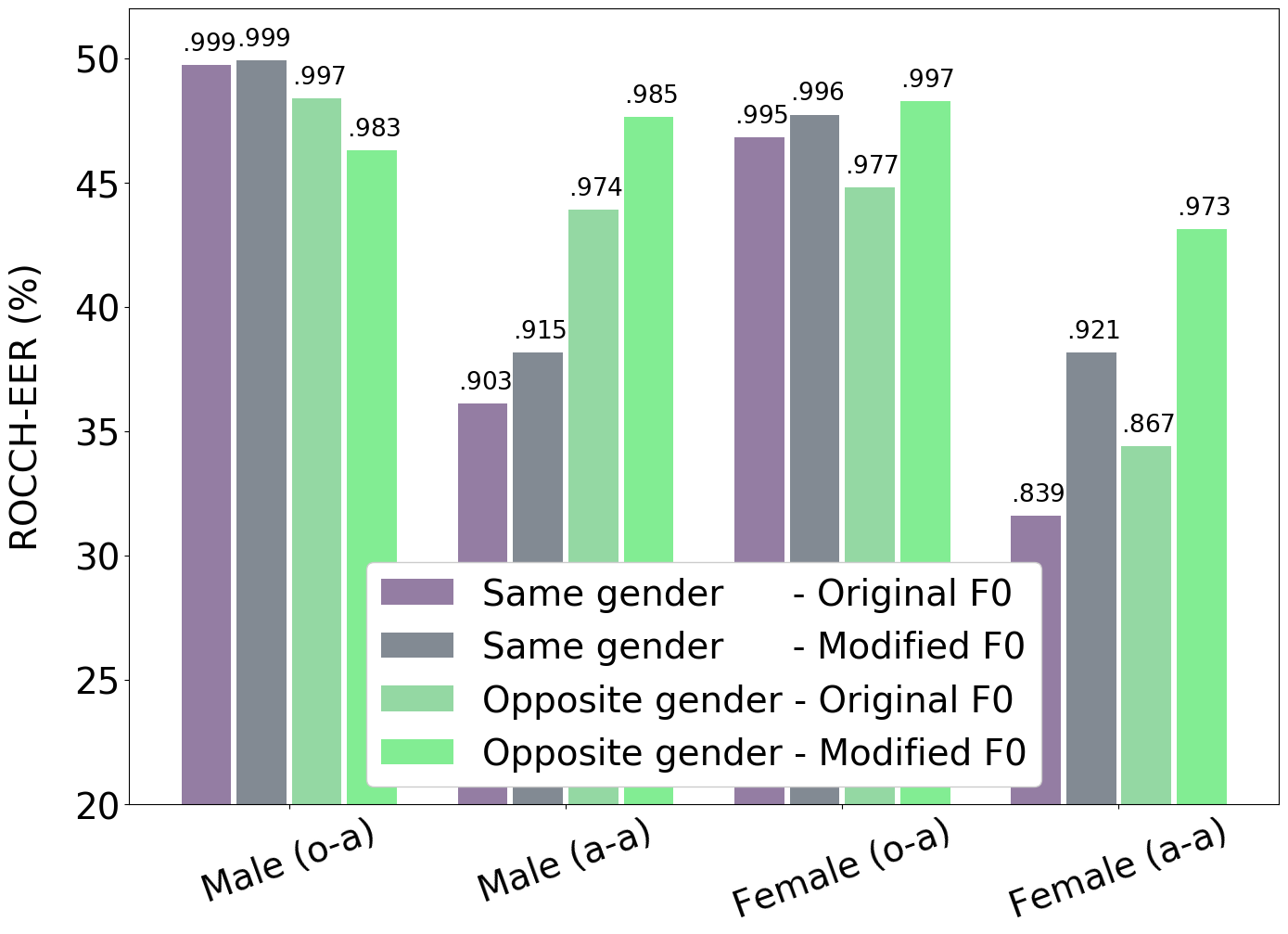}}
\caption{EER ($\%$) score obtained by the ASV evaluation system on Librispeech tests sets. The $\text{C}_{llr}^{min}$ score is displayed on the top of each bar. Multiple pipelines setups are reported for the gender selection and F0 modification. o – original, a – anonymized speech data for enrollment and trial parts. Entry \say{Same gender - Original F0} corresponds to the challenge baseline system.}
\label{fig:ASV-res}

\end{figure}

All results are compared to the VoicePrivacy baseline system. The pseudonymization pipeline with F0 modification contribution is publicly available\footnote{\url{https://github.com/deep-privacy/Voice-Privacy-Challenge-2020}}.
Figure \ref{fig:ASV-res} details the speaker linkability scores for \textbf{o}riginal to \textbf{a}nonymized (o-a) ASV tests, 
and for \textbf{a}nonymized to \textbf{a}nonymized (a-a) ASV tests in different gender selection and F0 modification setups. 
The \textbf{o}riginal to \textbf{a}nonymized test case helps to assess how capable systems are at modifying the original speech to make it sound like another speaker’s speech. 
As the system used to evaluate the linkability between \textbf{o}riginal and \textbf{a}nonymized speech is domain-dependent \cite{EvaluatingVoiceConversionbased2019}, and only trained on the original speech, it is thus of no surprise that the baseline provided in the challenge already shows great results.
As for the \textbf{a}nonymized to \textbf{a}nonymized test, enrolling the ASV system with anonymized data brings some speaker information in the process, although the pseudo-speaker x-vector is not exactly the same between random and trial utterances, because of the random part of the x-vector selection process (see Section on x-vector pseudonymization above).
Given this evaluation framework, our goal is to further degrade the linkability in both attacks models.
For each anonymization pipeline setups, the corresponding WER$_\%$ values are reported in Table \ref{tab:ASR-results}.

\subsection{Male linkability}
In the \textbf{o}riginal to \textbf{a}nonymized attack scenario (o-a in Figure \ref{fig:ASV-res}), we can observe that the proposed F0 modification doesn't affect the already good male un-linkability performance when compared to the challenge's baseline (\say{Same gender - Modified F0} compared to \say{Same gender - Original F0}). It appears that selecting an x-vector from the opposite gender without applying the F0 modification always degrades the pseudonymization un-linkability (\say{Opposite gender - Original F0} compared to \say{Same gender - Original F0}).
Applying the F0 modification together with the opposite gender x-vector selection doesn't improves performance.
This limitation might come from the x-vector selection algorithm, where the furthest speakers are selected to derive the pseudo-identity.

Regarding the \textbf{a}nonymized to \textbf{a}nonymized attack scenario (a-a in Figure \ref{fig:ASV-res}). Using the baseline anonymization setup, the attacker is able to re-identify the user at a much higher degree.
On their own, the F0 modification always improves compared to the baseline performance.
Jointly selecting the opposite gender and applying the F0 modification appears to be an excellent design choice against this attacker.

\subsection{Female linkability}
Contrary to the male results, the proposed F0 modification always improves the pseudonymization for female speaker in the \textbf{o}riginal to \textbf{a}nonymized attack scenario. This effect is observed regardless of the gender's x-vector selection (\say{Same gender - Modified F0} compared to \say{Same gender - Original F0} and \say{Opposite gender - Modified F0} compared to \say{Opposite gender - Original F0}).
Applying both the F0 modification and the opposite x-vector selection beats the baseline system.

The \textbf{a}nonymized to \textbf{a}nonymized attack scenario draws similar conclusions as for the male speaker. 
Jointly modifying gender for the x-vector selection and applying the F0 modification always improves pseudonymization. It is worth noting that female speakers are more sensitive to F0 modification than males. Meaning, the source's gender information plays a role in choosing the best anonymization procedure.

\subsection{Speech intelligibility}
\begin{table}[th]
\centering
\begin{tabular}{|c|c|c|c|c|c|}
\hline
 \textbf{Gender-selection} & \textbf{F0} & \textbf{Test} WER$_\%$\\
\hline \hline

  \multirow{2}{*}{Same} & Original & 6.73\\  \cline{2-3}
  & Modified & 6.92\\   \cline{1-3}
  \multirow{2}{*}{Opposite} & Original & 7.24\\  \cline{2-3}
  & Modified & 6.74\\ \cline{2-3}
 \hline
\end{tabular}
\caption{Speech recognition results in terms of WER$_\%$ for the LibriSpeech test set.}
\label{tab:ASR-results}
\end{table}
Across all experiments, the utility (Table \ref{tab:ASR-results}) is not tremendously affected by the gender x-vector selection, F0 modification, or the two modifications applied together. The high WER$_\%$ score (7.24) reported with the opposite x-vector gender selection, and no F0 modification might come from the fact that the ASR model used for the evaluation was trained on audiobooks data; and the fact that selecting opposite gender without modifying F0 might leads to some inconsistencies in the speech signal.


\section{Conclusions} \label{conclusion}
In this work, we proposed to alter the F0 paralinguistic information in an x-vector based speech pseudonymization system.
We evaluated this modification against the \textit{Opposite} and \textit{Same} gender x-vector target selection to obtain various anonymization setups.
We objectively evaluated the F0 modification using the VoicePrivacy 2020 challenge tools. The performance was assessed in terms of EER/$\text{C}_{llr}^{min}$ to measure privacy protection and WER$_\%$ to measure utility.
We observed that keeping the original F0 values retains some information about the original speaker. The experiments show that applying the F0 modification and selecting an x-vector from the \textit{Opposite} gender allows for better privacy protection against attackers who has access to the anonymization pipeline.
Our results also show that the performance of anonymization depends on the gender of the source. This raises the question of the importance of personalized modification in a privacy context.
In future work, we plan to subjectively evaluate the naturalness of the generated speech. We think the F0 modification helps to produce a more natural speech when an \textit{Opposite} gender's x-vector is selected. Because the F0 features will be coherent with the selected gender.

\section{Acknowledgments} 
This work was supported in part by the French National Research Agency under project DEEP-PRIVACY (ANR-18-CE23-0018) and Région Lorraine.
\bibliography{refs}

\begin{thebibliography}{20}
\providecommand{\natexlab}[1]{#1}
\providecommand{\url}[1]{\texttt{#1}}
\providecommand{\urlprefix}{URL }
\expandafter\ifx\csname urlstyle\endcsname\relax
  \providecommand{\doi}[1]{doi:\discretionary{}{}{}#1}\else
  \providecommand{\doi}{doi:\discretionary{}{}{}\begingroup
  \urlstyle{rm}\Url}\fi

\bibitem[{Bahmaninezhad, Zhang, and
  Hansen(2018)}]{F0_Bahmaninezhad2018ConvolutionalNN}
Bahmaninezhad, F.; Zhang, C.; and Hansen, J. H.~L. 2018.
\newblock Convolutional Neural Network Based Speaker De-Identification.
\newblock In \emph{Odyssey}.

\bibitem[{Brummer and Preez(2006)}]{cllr}
Brummer, N.; and Preez, J. 2006.
\newblock Application-independent evaluation of speaker detection.
\newblock \emph{Computer Speech \& Language} .

\bibitem[{Fang et~al.(2019)Fang, Wang, Yamagishi, Echizen, Todisco, Evans, and
  Bonastre}]{fangSpeakerAnonymizationUsing2019}
Fang, F.; Wang, X.; Yamagishi, J.; Echizen, I.; Todisco, M.; Evans, N.; and
  Bonastre, J.-F. 2019.
\newblock {Speaker Anonymization Using X-vector and Neural Waveform Models}.
\newblock In \emph{Proc. 10th ISCA Speech Synthesis Workshop}.

\bibitem[{Gussenhoven(2004)}]{PhonologyOfTone}
Gussenhoven, C. 2004.
\newblock \emph{Pitch in Language I: Stress and Intonation}.
\newblock Research Surveys in Linguistics. Cambridge University Press.

\bibitem[{Huang et~al.(2020)Huang, Luo, Hwang, Lo, Peng, Tsao, and
  Wang}]{f0_Huang2020UnsupervisedRD}
Huang, W.-C.; Luo, H.; Hwang, H.-T.; Lo, C.-C.; Peng, Y.-H.; Tsao, Y.; and
  Wang, H.-M. 2020.
\newblock Unsupervised Representation Disentanglement Using Cross Domain
  Features and Adversarial Learning in Variational Autoencoder Based Voice
  Conversion.
\newblock \emph{IEEE Transactions on Emerging Topics in Computational
  Intelligence} .

\bibitem[{Magari{\~n}os et~al.(2017)Magari{\~n}os, Lopez-Otero,
  Docio-Fernandez, Rodriguez-Banga, Erro, and
  Garcia-Mateo}]{magarinos2017reversible}
Magari{\~n}os, C.; Lopez-Otero, P.; Docio-Fernandez, L.; Rodriguez-Banga, E.;
  Erro, D.; and Garcia-Mateo, C. 2017.
\newblock Reversible speaker de-identification using pre-trained transformation
  functions.
\newblock \emph{Computer Speech \& Language} .

\bibitem[{McAdams(1984)}]{mcadams}
McAdams, S. 1984.
\newblock Spectral fusion, spectral parsing and the formation of the auditory
  image.
\newblock \emph{Ph. D. Thesis, Stanford} .

\bibitem[{Nautsch et~al.(2019)Nautsch, Jasserand, Kindt, Todisco, Trancoso, and
  Evans}]{nautschGDPRSpeechData2019}
Nautsch, A.; Jasserand, C.; Kindt, E.; Todisco, M.; Trancoso, I.; and Evans, N.
  2019.
\newblock {The GDPR \& Speech Data: Reflections of Legal and Technology
  Communities, First Steps Towards a Common Understanding}.
\newblock In \emph{Proc. Interspeech}.

\bibitem[{Parliament and Council(2016)}]{gdpr}
Parliament, E.; and Council. 2016.
\newblock Regulation (EU) 2016/679 of the European Parliament and of the
  Council of 27 April 2016 on the protection of natural persons with regard to
  the processing of personal data and on the free movement of such data, and
  repealing Directive 95/46/EC.
\newblock \emph{General Data Protection Regulation} .

\bibitem[{Povey et~al.(2011)Povey, Ghoshal, Boulianne, Burget, Glembek, Goel,
  Hannemann, Motlíček, Qian, Schwarz, Silovský, Stemmer, and
  Vesel}]{KaldiPovey}
Povey, D.; Ghoshal, A.; Boulianne, G.; Burget, L.; Glembek, O.; Goel, N.;
  Hannemann, M.; Motlíček, P.; Qian, Y.; Schwarz, P.; Silovský, J.; Stemmer,
  G.; and Vesel, K. 2011.
\newblock The Kaldi speech recognition toolkit.
\newblock \emph{IEEE Workshop on Automatic Speech Recognition and
  Understanding} .

\bibitem[{Qian et~al.(2020)Qian, Jin, Hasegawa-Johnson, and
  Mysore}]{F0_Qian2020F0ConsistentMN}
Qian, K.; Jin, Z.; Hasegawa-Johnson, M.; and Mysore, G.~J. 2020.
\newblock F0-Consistent Many-To-Many Non-Parallel Voice Conversion Via
  Conditional Autoencoder.
\newblock \emph{IEEE ICASSP} .

\bibitem[{Raj, Snyder, and Povey(2019)}]{prob_x-vector}
Raj, D.; Snyder, D.; and Povey, D. 2019.
\newblock Probing the Information Encoded in X-Vectors.
\newblock In \emph{IEEE ASRU}.

\bibitem[{Snyder et~al.(2018)Snyder, Garcia-Romero, Sell, Povey, and
  Khudanpur}]{snyder2018xvector}
Snyder, D.; Garcia-Romero, D.; Sell, G.; Povey, D.; and Khudanpur, S. 2018.
\newblock X-vectors: Robust DNN Embeddings for Speaker Recognition.
\newblock In \emph{IEEE ICASSP}.

\bibitem[{Srivastava et~al.(2020{\natexlab{a}})Srivastava, Tomashenko, Wang,
  Vincent, Yamagishi, Maouche, Bellet, and Tommasi}]{brij_vpc_design}
Srivastava, B. M.~L.; Tomashenko, N.; Wang, X.; Vincent, E.; Yamagishi, J.;
  Maouche, M.; Bellet, A.; and Tommasi, M. 2020{\natexlab{a}}.
\newblock Design Choices for X-vector Based Speaker Anonymization.
\newblock \emph{Proc. Interspeech} .

\bibitem[{Srivastava et~al.(2020{\natexlab{b}})Srivastava, {Vauquier},
  {Sahidullah}, {Bellet}, {Tommasi}, and
  {Vincent}}]{EvaluatingVoiceConversionbased2019}
Srivastava, B. M.~L.; {Vauquier}, N.; {Sahidullah}, M.; {Bellet}, A.;
  {Tommasi}, M.; and {Vincent}, E. 2020{\natexlab{b}}.
\newblock Evaluating Voice Conversion-Based Privacy Protection against Informed
  Attackers.
\newblock In \emph{IEEE ICASSP}.

\bibitem[{{Sun} et~al.(2016){Sun}, {Li}, {Wang}, {Kang}, and {Meng}}]{ppgs}
{Sun}, L.; {Li}, K.; {Wang}, H.; {Kang}, S.; and {Meng}, H. 2016.
\newblock Phonetic posteriorgrams for many-to-one voice conversion without
  parallel data training.
\newblock In \emph{IEEE ICME}.

\bibitem[{Tomashenko et~al.(2020)Tomashenko, Srivastava, Wang, Vincent,
  Nautsch, Yamagishi, Evans, Patino, Bonastre, Noé, and
  Todisco}]{tomashenkoVoicePrivacy2020Challenge}
Tomashenko, N.; Srivastava, B. M.~L.; Wang, X.; Vincent, E.; Nautsch, A.;
  Yamagishi, J.; Evans, N.; Patino, J.; Bonastre, J.-F.; Noé, P.-G.; and
  Todisco, M. 2020.
\newblock Introducing the {{VoicePrivacy Initiative}}.
\newblock \emph{Proc. Interspeech} .

\bibitem[{Ueda et~al.(2015)Ueda, Aihara, Takiguchi, and
  Ariki}]{F0_IndividualityPreservingSM}
Ueda, R.; Aihara, R.; Takiguchi, T.; and Ariki, Y. 2015.
\newblock Individuality-Preserving Spectrum Modification for Articulation
  Disorders Using Phone Selective Synthesis.
\newblock In \emph{Proc. Interspeech}.

\bibitem[{{Wang}, {Takaki}, and {Yamagishi}(2020)}]{nsf}
{Wang}, X.; {Takaki}, S.; and {Yamagishi}, J. 2020.
\newblock Neural Source-Filter Waveform Models for Statistical Parametric
  Speech Synthesis.
\newblock \emph{IEEE TASLP} .

\bibitem[{Zen et~al.(2015)Zen, Dang, Clark, Zhang, Weiss, Jia, Chen, and
  Wu}]{libritts}
Zen, H.; Dang, V.; Clark, R.; Zhang, Y.; Weiss, R.~J.; Jia, Y.; Chen, Z.; and
  Wu, Y. 2015.
\newblock LibriTTS: A Corpus Derived from LibriSpeech for Text-to-Speech.
\newblock In \emph{Proc. Interspeech}.

\end{thebibliography}

\end{document}